\documentclass[11pt,twoside]{article}
\usepackage{asp2010}
\usepackage{graphicx}

\resetcounters

\begin{document}

\title{Population Synthesis of Normal Radio and Gamma-ray Pulsars Using Markov Chain Monte Carlo Techniques}
\author{Peter L. Gonthier$^1$, Caleb D. Billman$^1$, and Alice K. Harding$^2$
\affil{$^1$Department of Physics, Hope College, 27 Graves Place, Holland, Michigan 49423}
\affil{$^2$NASA Goddard Space Flight Center, 8800 Greenbelt Rd., Greenbelt, MD., 20771}}

\begin{abstract}
We present preliminary results of a pulsar population synthesis of normal pulsars from the Galactic disk using a Markov Chain Monte Carlo method to better understand the parameter space of the assumed model. We use the Kuiper test, similar to the Kolmogorov-Smirnov test, to compare the cumulative distributions of chosen observables of detected radio pulsars with those simulated for various parameters.  Our code simulates pulsars at birth using Monte Carlo techniques and evolves them to the present assuming initial spatial, kick velocity, magnetic field, and period distributions. Pulsars are spun down to the present, given radio and gamma-ray emission characteristics, filtered through ten selected radio surveys, and a {\it Fermi} all-sky threshold map.  Each chain begins with a different random seed and searches a ten-dimensional parameter space for regions of high probability for a total of one thousand different simulations before ending.  The code investigates both the``large world" as well as the ``small world" of the parameter space.  We apply the K-means clustering algorithm to verify if the chains reveal a single or multiple regions of significance.  The outcome of the combined set of chains is the weighted average and deviation of each of the ten parameters describing the model.  While the model reproduces reasonably well the detected distributions of normal radio pulsars, it does not replicate the predicted detected $\dot P - P$ distribution of {\it Fermi} pulsars.  The simulations do not produce sufficient numbers of young, high-$\dot E$ pulsars in the Galactic plane.
\end{abstract}

Simulating the characteristics of pulsars calls for the use of computer intensive Monte Carlo techniques that model various important distributions with many parameters that are not known a priori.  In this study, we explore the parameter space using Markov Chain Monte Carlo (MCMC) techniques to understand regions in the parameter space that produce distributions that agree best with those detected.  First we establish a birth model that defines the present-day spatial distribution of neutron stars in the Galaxy.  The code seeds the Galaxy with neutron stars (NS) using an initial exponential scale height of 50 pc, a radial distribution \citep{Isab, Bania} and a supernova kick velocity distribution \citep{Hobbs}. The code evolves the NSs in the Galactic potential \citep{Pac} from their birth location to the present day.  This group of evolved NSs has no free parameters and represents the present-day spatial model of NSs with the Galaxy that we use in this study.  Of course, their are other options with different birth spatial distributions, like within the spiral arms, and different supernova kick velocity distributions, which we hope to explore in the future.

The MCMC code takes this present-day model of NSs, assigns an initial period $P_o$ and magnetic field $B$ from a 2D log-normal in $B$ and Gaussian in $P_o$ distribution described with 5 free parameters that characterize the means, widths and the correlation parameter.  The code then spins down the NSs from their initial period to their present $P$ and $\dot P$ using a braking index model with the index being a free parameter; this model is similar in nature to the one used by \cite{Ridley} without the dependence on the inclination angle.  Another free parameter describes the log of the alignment constant of the exponential inclination angle model.  Three more free parameters are used to describe the radio luminosity as power laws in $P$ and $\dot P$ for a total of ten free parameters.  The radio beam geometry \citep{HGG07} has no fitted parameters as well as the $\gamma$-ray slot gap beam and luminosity \citep{Pier} model.  The favorable step taken by the MCMC in the ten-dimensional parameter space is based on the sum of the maximum positive and negative differences in the cumulative distributions between the detected and simulated distributions of $P$, $\dot P$, $DM$ and $S_{1400}$, where $DM$ is the dispersion measure, and $S_{1400}$ is the radio luminosity at 1400 MHz.  This test is similar to the Kolmogorov-Smirnov test and is known as the Kuiper test \citep{Press}.  We ran 38 MCMC chains each for 1000 simulations.  The ``best" choice set of parameters at the end of 1000 simulations for each chain is averaged being weighted by the final minima of the differences in the cumulative distributions.  The average set of ten free parameters is then used to run a complete simulation whose results we show below.

In Figure 1, we present very preliminary results with 1D histograms comparing the indicated pulsar characteristics of simulated (open histograms) and detected (solid histograms) NPs in a select group of ten radio surveys.  What is surprising is that the \begin{figure}[!ht]\begin{center}
\includegraphics[scale = 0.45]{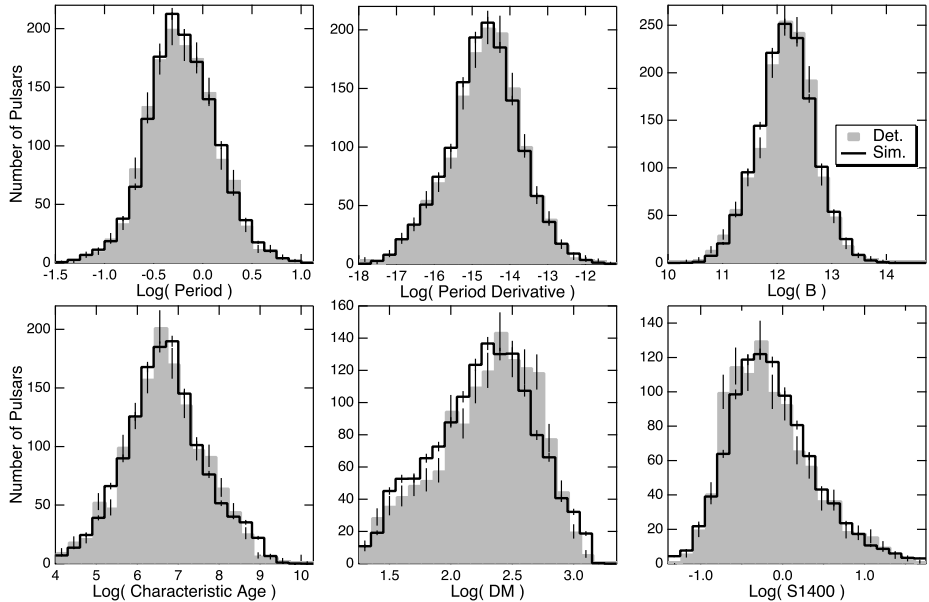}
\end{center}
\caption{Histograms of the indicated characteristics of detected (solid) and simulated (open) normal pulsars.  The error bars represent statistical uncertainties.  To generate the simulated histograms, the code was run for ten times the number of observed pulsars.}
\end{figure}

\noindent MCMC simulations preferred a strong negative or anti-correlation of $-0.5$ between the initial $\log (B)$ and $P_o$, which suggests that NSs born with lower magnetic fields have larger birth periods.  The values of the other free parameters are more expected where $\mu_{\log B} = 12.7$, $\sigma_{\log B} = 0.7$, $\mu_{P_o}=0.26$, $\sigma_{P_o} = 0.21$, a braking index of 2.8, the $\log$ of the alignment constant of 7.0, the period index of -2.1 and period derivative index of 0.5 of the radio luminosity.

NASA's Fermi Gamma-Ray Space Telescope, ({\it Fermi}), has opened a new window in pulsar $\gamma$-ray astronomy by discovering more that 100 pulsars that include over 40 millisecond (MSP) and over 60 normal (NP) pulsars, superseding the $\gamma$-ray pulsar database of six provided by its predecessor, the EGRET instrument aboard the Compton Gamma-Ray Observatory.  The Second Pulsar Catalogue will soon be available this summer.  While we did not use a comparative statistic between simulated and detected $\gamma$-ray pulsars, we did simulate {\it Fermi} pulsars within the slot gap model \citep{Pier}.  In this study, we focus on a group of 46 public {\it Fermi} pulsars detected during an approximate period of two years.  

In Figure 2, we present the $\dot P - P$ diagram of the detected (left panel) and simulated (right panel) normal radio (dots) and $\gamma$-ray (hourglasses) pulsars. While the simulated radio distribution reproduces well the funnel-shaped distribution observed in the detected radio pulsars, the stark difference occurs with simulated {\it Fermi} pulsars, which lack the large number of young, high-$\dot E$ detected {\it Fermi} pulsars.  Younger and higher $\dot E$ pulsars are more $\gamma$-ray luminous, yet are not seen in the simulated distributions.  
\begin{figure}[!ht]\begin{center}
\includegraphics[scale = 0.4]{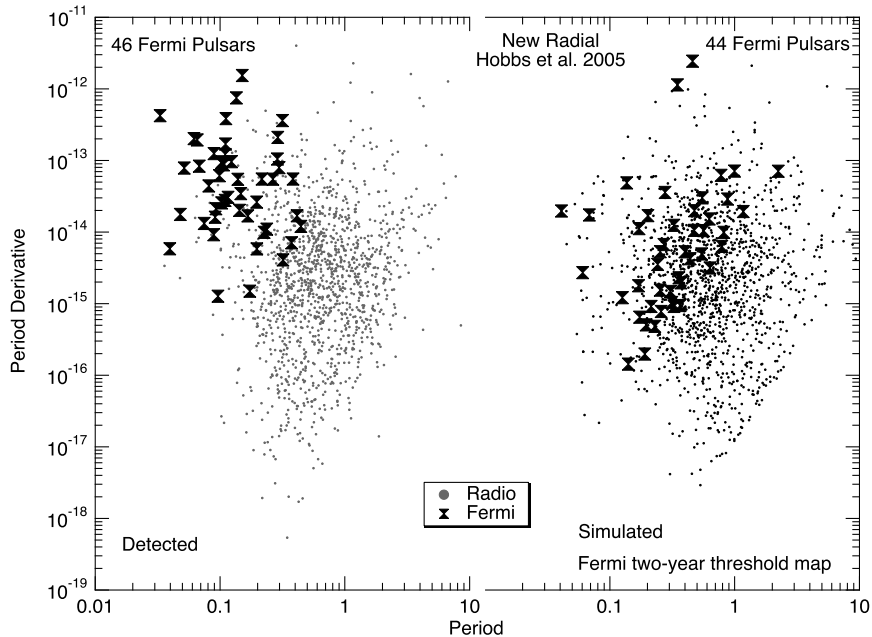}
\end{center}
\caption{Period derivative - period diagram for normal pulsars detected (left) and simulated (right).  The small solid dots represent radio pulsars and the larger hourglass symbols represent {\it Fermi} pulsars.}
\end{figure}
\noindent If the $\gamma$-ray luminosity is increased, more simulated {\it Fermi} pulsars are seen in the same region of the $\dot P - P$ diagram suggesting that the pulsars are simply not present in the present-day spatial distribution of NS.  For more comments on this discrepancy, see \citet*{Pier}.

In summary, our main focus of this study is the new implementation of MCMC to explore the ten-dimensional free parameter space of our model to find regions of best agreement in the select detected of radio pulsar characteristics of $P$, $\dot P$, $DM$ and $S_{1400}$.  While the use of our implementation of MCMC is not very efficient in finding the best regions, we wanted to explore the ten-dimensional parameter space as much as possible.  As a result, we find an unexpected anti-correlation between the birth $\log B$ and $P_o$, suggesting that lower field pulsars are born with larger initial periods.  Perhaps this anti-correlation is what might be required to account for the group of central compact objects \citep{Kas10, Halp10, Mig09}.  The simulation also supports an inclination angle alignment with a $\log \tau_\alpha = 7.0$, smaller than the one obtained in recent studies \citep{WJ08} and \citep{MGR11}. In the interpulse study of  \citep{MGR11}, for example, an interpulse from the opposite pole is observed in a radio pulsar with a characteristic age of 100 Myr.  This particular pulsar, J1915+1410, is nearly an orthogonal rotor as required to detect the interpulse from the opposite pole, and if the inclination alignment models are correct, this pulsar must be born essentially with its period of 0.297 s and with a field of $\sim\log B=11$.

As indicated these results are very preliminary, and the effort is in progress. We hope to implement the next generation of population synthesis in the near future to better understand the confidence regions of our parameter space and to better understand what improvements can be deployed in our simulations, particularly with regard to account for the detected {\it Fermi} pulsars, which seem to suggest a required much larger group of young pulsars than those suggested by the simulations of normal radio pulsars.

\acknowledgements We are also grateful for the generous support of the National Science Foundation (grants AST-1009731 and REU PHY/DMR-1004811),  the NASA Astrophysics Theory Program through grant  NNX09AQ71G and the Michigan Space Grant Consortium.

\bibliography{GonthierTalk}

\end{document}